\documentclass[12pt]{article}
\usepackage{amstext}
\def\be{\begin{eqnarray}}
\def\ee{\end{eqnarray}}
\def\ba{\begin{array}} 
\def\ea{\end{array}}

\begin{document}
\begin{center}
{\large \bf   	
Least action principle for Lorentz force in  dilaton-Maxwell electrodynamics} 
\end{center}

\begin{center}

I.P. Denisova$^{*,\S}$  and O.V. Kechkin$^{*,\dag,\ddag,\star}$  

\end{center}

\begin{center}
{\it \qquad $*$ Moscow Aviation Institute (National Research University),\\
Moscow 125993, Russia,\\
$\dag$ D.V. Skobeltsyn Institute of Nuclear Physics, \\ M. V. Lomonosov Moscow State University,  Moscow 119234, Russia,\\
\quad $\ddag$ Faculty of Physics, M. V. Lomonosov Moscow State University,  \\
Moscow 119991, Russia\\

$\S$ pm@mati.ru  \\ 
$\star$ kechkin@srd.sinp.msu.ru } 
\end{center}

\vskip 0.5cm

\begin{center}
{\Large\bf Abstract}
\end{center}

\noindent 
The least action principle is established for the dynamics of a test particle in a dilaton-Maxwell background. These dynamics and background are invariant under the action of the dilatation transformation; explicit form of the corresponding generalization of the Lorentz force is established for the considered model. On a stationary background, we have found the integral of motion of the energy type. This integral is used to resolve the radial dynamics of test particles in a spherically symmetric electrostatic background. 

\vskip 0.5cm



\section*{Introduction}

A dilaton generalization of Maxwell's electrodynamics (dilaton-Maxwell electrodynamics, DME) is predicted by various Grand unification theories \cite{KK-1}--\cite{SS}. The action for this theory has the following form: 
\be\label{17-1-(-5)} {\cal S}_{b}=- \int d^4x\, \left(\frac{1}{4}e^{-2\alpha\phi}F_{\mu\nu}F^{\mu\nu}+2\sigma \partial_{\mu}\phi\partial^{\mu}\phi\right),\ee
where $F_{\mu\nu}=\partial_{\mu}A_{\nu}-\partial_{\nu}A_{\mu}$ ($\mu,\,\nu,\dots=0,\,\dots,\,3$), whereas  $\alpha$ is the coupling constant between the dilaton and the Maxwell fields $\phi=\phi\left(x^{\nu}\right)$ and $A_{\mu}=A_{\mu}\left(x^{\nu}\right)$, respectively (alternative nonlinear generalizations of classical electrodynamics can be found in  \cite{NED-1}--\cite{NED-3}). DME is considered on a flat four-dimensional background, the indices are raised and lowered using the Minkowski metric $\eta^{\mu\nu}=\eta_{\mu\nu}={\rm diag}\left(1,\,-1,\,-1,\,-1\right)$. Then, $\sigma=\pm 1$:  it is important to include in our consideration of cases with usual and phantom dilaton field. 

A characteristic property of the theory (\ref{17-1-(-5)}) is its invariance in respect to the dilatation transformation 
\be\label{17-1-(-4)} \phi\rightarrow \phi+\Lambda \qquad A_{\mu}\rightarrow e^{\alpha\Lambda}A_{\mu},\ee
(with $x^{\mu}= {\rm inv}$), where $\Lambda={\rm const}$  is an arbitrary real parameter. In \cite{DME-1} it was shown that this symmetry extends to the symmetry group $SL(2,\,R)$ in the electro(magneto)static case in which DME is dual to stationary General Relativity in vacuum (this correspondence between the theories is achieved at $\sigma =+1$). Then, the dilatation-invariant dynamics of a test particle in the DME background fields was developed in \cite{DME-2}. There it has been postulated that the trajectory $x^{\mu}=x^{\mu}\left(s\right)$ of test particle of mass m and charge n is determined by the modified Lorentz force as
\be\label{17-1-(-3)} \frac{du^{\mu}}{ds}=\frac{q}{m}\,e^{-\alpha\phi}F^{\mu\nu}u_{\nu},\ee
where $u^{\mu}=dx^{\mu}/ds$ is a four-velocity, and $ds^2=\eta_{\mu\nu}dx^{\mu}dx^{\nu}$. It is seen that Eq. (\ref{17-1-(-3)}) indeed is invariant under the transformation (\ref{17-1-(-4)}).

The generalization (\ref{17-1-(-3)}) of the conventional expression of classical electrodynamics satisfies the correspondence principle; it is based on algebraic arguments, and the analytical simplicity of the resulting dynamic scheme defined by Eqs. (\ref{17-1-(-5)}) and (\ref{17-1-(-3)}). However, Eq. (\ref{17-1-(-3)}) is not the equation of Euler-Lagrange type for any action, while the dynamics of DME background fields satisfy the principle of least action (see Eq. (\ref{17-1-(-5)})). In addition, it seems natural to have extra gradient term for the dilaton field (that is, the term $\sim\partial_{\nu}\phi$), which is absent in the present form of the modified Lorentz force (\ref{17-1-(-3)}).

In this paper, we propose a solution to these problems using the geometrical approach based on the Kaluza-Klein theory, which leads to the action of the form (\ref{17-1-(-5)}) in the special case $\alpha=\sqrt 3$ \cite{KK-1}. Namely, parameterizing five-dimensional line element as
\be\label{17-1-(-2)} ds_{(5)}^2=g^{(5)}_{AB}dx^Adx^B=e^{\frac{2\phi}{\sqrt 3}}ds_{(4)}^2+\sigma e^{-\frac{4\phi}{\sqrt 3}}\left(dx^4+A_{\mu}dx^{\mu}\right)^2,\ee
where $A,\,B=0,\,\dots,\,4$ and $ds_{(4)}^2=g^{(4)}_{\mu\nu}dx^{\mu}dx^{\nu}$,
one obtains the DME action (\ref{17-1-(-5)}) with $\alpha=\sqrt 3$ from the Kaluza-Klein theory one ${\cal S}_{KK}\sim \int \, d^5 x \sqrt{|{\rm det} \,g^{(5)}|}\,R^{(5)}$  after imposing the restrictions  
\be\label{17-1-(-1)} \partial_4\phi=\partial_4A_{\mu}=0\ee
and $g^{(4)}_{\mu\nu}=\eta_{\mu\nu}$ (where $g^{(5)}$ and $R^{(5)}$ are the determinant and curvature scalar, respectively, corresponding to  the five-metric $g^{(5)}_{AB}$). Below we formulate the least action principle, based on which we obtain the explicit form of the modified Lorentz force in the theory with an arbitrary value of the dilaton-Maxwell coupling $\alpha$. Our approach is inspired by the ideas and relations of the Kaluza-Klein theory (in fact, we develop a suitable generalization to a purely geometric scheme (\ref{17-1-(-2)})--(\ref{17-1-(-1)})). 
 
\section{Modified Lorentz force}

Let us take the action of a test particle in the following form:  
\be\label{17-1-1}
{\cal S}_p=- m\int d\lambda {\cal L}_p,\ee
where 
\be\label{17-1-2}
{\cal L}_p=\left[e^{2a\phi}\eta_{\mu\nu}\dot x^{\mu}\dot x^{\nu}+\sigma e^{2b\phi}\left(\dot x^4+A_{\mu}\dot x^{\mu}\right)^2\right]^{\frac{1}{2}},
\ee
and $a,\,b ={\rm const}$ (for example, $a=1/\sqrt 3$,\, $b=-2/\sqrt 3$ for the case of Kaluza-Klein theory, see Eq. (\ref{17-1-(-2)})). This action determines the trajectory $x^A=x^A(\lambda)$ in terms of the parameter $\lambda$ (here $\dot x^A=dx^A/d\lambda$). It is easy to see that Eqs. (\ref{17-1-(-4)}), being complemented by the map
\be\label{17-1-3}
x^4\rightarrow e^{\alpha\Lambda}x^4,
\ee
realise the dilatation symmetry in the dynamics defined by Eqs. (\ref{17-1-1}) and (\ref{17-1-2}) if the constraint 
\be\label{17-1-4}
a=b+\alpha
\ee
are imposed on the free parameters $a$ and $b$ of this model. Indeed, (\ref{17-1-(-4)}) and (\ref{17-1-3}) generate a simple rescaling of the Lagrangian (\ref{17-1-2}) in this case: 
\be\label{17-1-5}
{\cal L}_p\rightarrow {\it Const} \cdot {\cal L}_p
\ee
(with ${\it Const}=e^{a\Lambda}$), so that the resulting dynamics of the test particle is invariant under this extended dilatation transformation. Note that the restriction (\ref{17-1-4}) becomes the identity ($1/\sqrt 3=-2/\sqrt 3+\sqrt 3$) in the Kaluza-Klein theory case.

Now let us study the motion equations for the test particle defined by the Kaluza-Klein type action (\ref{17-1-1})--(\ref{17-1-2}) and  constraints (\ref{17-1-(-1)}). The $x^4$--equation
$d\left[e^{2b\phi}\left(\dot x^4+A_{\mu}\dot x^{\mu}\right)/{\cal L}_p\right]/d\lambda=0$
can be integrated; the result reads:
\be\label{17-1-6}
\frac{e^{2b\phi}}{{\cal L}_p}\left(\dot x^4+A_{\mu}\dot x^{\mu}\right)=\kappa,
\ee
where $\kappa={\rm const}$. From Eqs. (\ref{17-1-(-4)}) and (\ref{17-1-3}) it follows that 
\be\label{17-1-8}
\kappa\rightarrow e^{b\Lambda}\kappa.
\ee
Thus, the parameter $\kappa$ is an invariant of the dilatation transformation only in the special case $b=0$. Then, using Eqs. (\ref{17-1-2}) and (\ref{17-1-6}), one concludes that the differential relation
\be\label{17-1-9}
{\cal L}_pd\lambda=\frac{e^{a\phi}ds}{\left(1-\sigma\kappa^2e^{-2b\phi}
\right)^{\frac{1}{2}}}
\ee
is performed on the considered trajectory. Finally, using Eq. (\ref{17-1-9}), the $x^{\mu}$--equation can be reduced to the form of the Cauchy problem for the test particle in DME background fields:
\be\label{17-1-10}
\frac{du^{\mu}}{ds}={\cal P}F^{\mu\nu}u_{\nu}+{\cal Q}\left(\eta^{\mu\nu}-u^{\mu}u^{\nu}\right)\partial_{\nu}\phi,
\ee
where
\be\label{17-1-11-1}
{\cal P}=\frac{\sigma\kappa e^{-a\phi}}{\left(1-\sigma\kappa^2e^{-2b\phi}
\right)^{\frac{1}{2}}},\qquad {\cal Q}=a+\frac{\sigma b\kappa^2e^{-2b\phi}}{1-\sigma\kappa^2e^{-2b\phi}}.
\ee
Note, that the right side of Eq. (\ref{17-1-10}) defines a modified Lorentz force in this theory. Of course, the dynamics of the test particle is invariant with respect to dilatation transformation (\ref{17-1-(-4)}) 
(actually, ${\cal P}\rightarrow e^{\alpha\Lambda}{\cal P}$,\, ${\cal Q}\rightarrow {\cal Q}$ in view of Eqs. (\ref{17-1-4}) and (\ref{17-1-8})). Then, the $\mu=k$ part of Eq. (\ref{17-1-10}) (with $k=1,\,2,\,3$) defines the three-dimensional form for the Lorentz force. The result of the calculation is as follows:
\be\label{17-1-11-2}
\frac{d\vec u}{dt}={\cal P}\left(\vec E+\vec v\times \vec H\right)
-{\cal Q}\left[\gamma^{-1}\nabla\phi+\gamma\left(\partial_t\phi+\vec v\nabla\phi\right)\vec v\right],\ee
where $\vec v=d\vec r/dt$, $\vec u=\gamma\vec v$ and $\gamma=1/\sqrt{1-v^2}$, while $\vec E$ and $\vec H$ are electric and magnetic fields, respectively, which are defined as in the conventional  electrodynamics of Maxwell.

It is clear that for the DME background with asymptotically trivial dilaton field (that is, for DME solutions with $\phi\left(r\rightarrow +\infty\right)=0$) one gets the following interpretation of the constant parameters of the theory:
\be\label{17-1-12}
{\cal P}\left(r\rightarrow +\infty\right)=\frac{\sigma\kappa }{\left(1-\sigma\kappa^2
\right)^{\frac{1}{2}}}=\frac{q}{m},
\ee
where $q$ and $m$ are the electric charge and the mass of the test particle, respectively.
It is seen that the most similar form of Eqs. (\ref{17-1-(-3)}) and 
(\ref{17-1-10}) is achieved at $b=0$; in this case 
\be\label{17-1-13}
\frac{du^{\mu}}{ds}=\frac{q}{m}e^{-\alpha\phi}F^{\mu\nu}u_{\nu}+\alpha\left(\eta^{\mu\nu}-u^{\mu}u^{\nu}\right)\partial_{\nu}\phi.
\ee
Thus, the purely dilaton part of the modified Lorentz force (that is, the contribution to this force $\sim\partial_{\nu}\phi$) is predicted by the principle of least action for test particles for EMD with $\alpha\neq 0$.
This contribution does not disappear for DME backgrounds with trivial electromagnetic sector.  

\section{Modified integral of energy}

Equations (\ref{17-1-10}) and (\ref{17-1-11-1}) completely define the dynamics of the test particle on the EMD background. Both the background and the dynamics are invariant under the action of the dilatation transformation (\ref{17-1-(-4)}) (with the additional condition (\ref{17-1-4}) in the case of the dynamics of the test particle). It is important to emphasize that the resulting dynamic scheme, being highly nonlinear, is based on the fundamental principle of least action. Namely, the EMD background is associated with the action (\ref{17-1-(-5)}), while the dynamics of the test particle is derived from the action (\ref{17-1-1})--(\ref{17-1-2}). In this section we construct the integral of motion for the dynamics of a test particle in a stationary DME background,  that is, in the special case of
\be\label{17-1-14}
\partial_t\phi=\partial_t A_{\mu}=0.
\ee
The corresponding result is obtained for arbitrary values of the parameters $a$ and $b$ (in particular, the integral of motion is constructed for a dilatation-invariant version of the theory specified by Eq. (\ref{17-1-4})).

Namely, let us consider $\mu=0$ part of Eq. (\ref{17-1-10}). It is easy to see that this relation can be rewritten in the following form:   
\be\label{17-1-15}
{\cal P}^{-1}\frac{du^0}{dt}+\left({\cal P}^{-1}{\cal Q}\,\frac{d\phi}{dt}\right) u^0+\frac{dA^0}{dt}=0, 
\ee
where $dA^0/dt= v^k\partial_kA^0$ is the total time derivative of the electrostatic potential in the stationary case. Then,
using the explicit form (\ref{17-1-11-1}) for quantities ${\cal P}$ and ${\cal Q}$, it is not difficult to prove that the remarkable relation 
\be\label{17-1-16}
\frac{d{\cal P}^{-1}}{d\phi}={\cal P}^{-1}{\cal Q}
\ee
is valid. Thus, from Eqs. (\ref{17-1-15}) and (\ref{17-1-16}) it follows that
$d\left({\cal P}^{-1}u^0+A^0\right)/dt=0$, that is, we have the integral of motion
\be\label{17-1-17}
{\cal P}^{-1}u^0+A^0=\tilde {\cal E}
\ee
(where $\tilde {\cal E}={\rm const}$) in the considered stationary case. It is clear that this quantity is of the energy type: a simple analysis shows that in the case of asymptotically trivial DME background (with $\phi\left(r\rightarrow +\infty\right)=A^0\left(r\rightarrow +\infty\right)=0$) one has
\be\label{17-1-18}
\tilde {\cal E}=\frac{{\cal E}}{q},
\ee
where ${\cal E}=mu^0\left(r\rightarrow +\infty\right)$ is the relativistic energy of a particle at the spatial infinity (that is, the corresponding sum of its kinetic and rest energies). Moreover, using Eqs. (\ref{17-1-12}) and (\ref{17-1-18}), one can rewrite Eq. (\ref{17-1-17}) in a more familiar form ${\cal P}^{-1}p^0+qA^0={\cal E}$, where $p^0=mu^0$ and $qA^0$ are relativistic and electrostatic energy of the particle, respectively. Thus, in this case, the conserved quantity ${\cal E}$ is a modification of the conventional total energy of a test particle whose dynamics takes place on the background of asymptotically trivial fields of DME.

\section{Radial dynamics on central background}

Now let us consider the radial dynamics of a test particle in a central electrostatic DME background, that is, let's take a consistent ansatz $\vec A= 0$, $A^0=A^0(r)$ and $\phi=\phi(r)$ for the background fields of the system. In this special case $\vec E=-A^{0}_{,\,r}\vec e_r$ and $\nabla\phi=\phi_{,\,r}\vec e_r$, where $\vec e_r=\vec r/r$ denotes the radial unit vector; and $\vec u=u\vec e_r$, where $u=u(r)$ for the considered radial dynamics. It is easy to prove that Eq. (\ref{17-1-11-2}) is equivalent to the `radial'\, equation of motion
\be\label{17-1-19}
\frac{du}{dt}=-\left({\cal P}A^{0}_{,\,r}+\gamma{\cal Q}\phi_{,\,r}\right);
\ee
the resulting dynamic system is free from any additional restrictions. 
Then, taking into account that $\gamma=u^0$, and using Eq. (\ref{17-1-17}) for the exclusion of $\gamma$ from Eq. (\ref{17-1-19}), eventually come to the equation
\be\label{17-1-20}
\frac{du}{dt}=-\frac{dV}{dr}
\ee
with
\be\label{17-1-21}
V={\cal P}\left(A^0-\tilde {\cal E}\right).
\ee
Thus, this dynamics has a potential nature. The effective potential $V=V(r,\,\tilde {\cal E})$  is determined by Eq. (\ref{17-1-21}); it depends on the parameter $\tilde {\cal E}$,  which can be expressed in terms of the initial data of the problem.  
In fact, this dynamic is solvable by help of its first integral (\ref{17-1-17}). Actually, taking into account that Eq. (\ref{17-1-17}) is equivalent to the relation  
\be\label{17-1-21-2} 
u^0+V=0
\ee 
where $u^0=\gamma=1/\sqrt{1-v^2}$ and $v=dr/dt$ for radial motion, we obtain the solution
\be\label{17-1-22}
t=t_0\pm\int_{r_0}^{r}\frac{dr'}{\left[1-V^{-2}\left(r'\right)\right]^{\frac{1}{2}}}
\ee
for the trajectory $r=r(t)$ of test particle with initial data $r_0=r(t_0)$, where we choose `$+$' (`$-$') for $r\geq r_0$ ($r\leq r_0$).

Then, in \cite{DME-1} it was studied a static background of DME with $\sigma=+1$ (it has been shown that this choice of the sign leads to a remarkable duality between static DME and stationary General Relativity in vacuum). Let's fix this type of DME; in \cite{DME-1} it was shown that a general spherically symmetric solution of the DME equations with trivial spatial asymptotics reads:  
\be\label{17-1-23}
A^0=\frac{Q_e\sinh \left(\frac{r_0}{r}\right)}{r_0\Pi}, \qquad e^{-\alpha\phi}=\Pi=\cosh\left(\frac{r_0}{r}\right)-\frac{\alpha Q_d}{r_0}\sinh \left(\frac{r_0}{r}\right),
\ee
where 
\be\label{17-1-24}
r_0=\frac{\alpha\sqrt{Q_e^2+4Q_d^2}}{2}.
\ee
Here $Q_e$ and $Q_d$ are arbitrary real constants which can be interpreted as the electric and dilaton charges, respectively (indeed, from Eq. (\ref{17-1-17}) it follows that $A^0\rightarrow Q_e/r$ and $\phi\rightarrow Q_d/r$, if $r\rightarrow\infty$). For example, in the analytically simplest case with $b=0$ (which is closest to the model considered in \cite{DME-2}), one obtains the following explicit form of the potential energy $U=mV$
\be\label{17-1-25}
U\left(r\right)=\frac{Q_eq+\alpha Q_d{\cal E}}{r_0}\sinh\left(\frac{r_0}{r}\right)-{\cal E}\cosh \left(\frac{r_0}{r}\right)
\ee
after substitution of Eqs. (\ref{17-1-18}) and (\ref{17-1-23}) into Eq. (\ref{17-1-21}). It is seen that $U\left(r\right)\rightarrow \left(Q_eq+\alpha Q_d{\cal E}\right)/r$ at $r\rightarrow\infty$; thus, the potential energy demonstrates the Coulomb behavior at spatial infinity,  but the product of the charges $Q_eq$ in it is modified by a `dilaton--energy'\, contribution $\alpha Q_d{\cal E}$. This contribution describes additional interaction in the system, which is characterized by relativistic energy of the test particle and the scalar charge of the source. 

Then, substituting Eq. (\ref{17-1-25}) into (\ref{17-1-22}) (as $V=U/m$), we obtain the solution in quadratures for the dynamics of a test particle in the central background. As usual, the classical turning points of this dynamics correspond to the vanishing (relativistic) kinetic energy $T=m(u^0-1)$ of the test particle; in our case this condition is equivalent to the relation $U=-m$ due to Eq. (\ref{17-1-21-2}). We hope to further study the respective effects in the upcoming publication.
  
Note finally that a potential in the theory with arbitrary value of the parameter $b$ can be obtained from the potential in the theory with $b=0$ by means of the mapping 
\be\label{17-1-26}
V\rightarrow \frac{\Pi^{\beta}}{\left[1+\sigma\left(\frac{q}{m}\right)^2\left(1-\Pi^{2\beta}\right)\right]}\,V,
\ee
where we put $b=\alpha\beta$ (this result immediately follows from Eqs. 
(\ref{17-1-21}), (\ref{17-1-11-1}), (\ref{17-1-4}), (\ref{17-1-12}) and (\ref{17-1-23})).
Thus, the most general form of the potential for the radial dynamics of a test particle on the background of DME with a given value of the dilaton-Maxwell coupling $\alpha$ contains the free parameter $\beta$ which can not be fixed by the underlying dilatation symmetry (\ref{17-1-(-4)})
(for example, $\beta=-2/3$ in the Kaluza-Klein theory case).

\section*{Conclusion}

In this article on the basis of the principle of least action, we have obtained the explicit form for the dilaton generalization of the Lorentz force. The result is given by Eqs. (\ref{17-1-10}) and (\ref{17-1-11-1}); in addition, the relation (\ref{17-1-4}) should be taken into account. By construction, the resulting dynamic scheme is symmetric with respect to dilatation transformation (\ref{17-1-(-4)}), (\ref{17-1-8}) (DME background (\ref{17-1-(-5)}) possesses this symmetry `initially'). It is interesting to note that the principle of least action predicts the presence of a dilaton component of the generalized Lorentz force which does not disappear in the trivial case of the electromagnetic field in contrast to the alternative version (\ref{17-1-(-3)}) obtained on the basis of purely algebraic arguments.  

It is clear that dilatation symmetry is a global symmetry, and that the consideration of the class of asymptotically trivial DME backgrounds (with $\phi (r\rightarrow\infty)=0$) is related to the fixation of the dilatation gauge. Thus, we are dealing here with a dilatation symmetry breaking, and this creates the illusory problem with the application of transformation (\ref{17-1-8}) to relation (\ref{17-1-12}). In fact, more careful analysis shows that the problem is completely absent, because it is necessary to put $\kappa={\rm inv}$ in this relation.  
Actually, let us rewrite the first term in Eq. (\ref{17-1-11-1}) as follows:   
\be\label{17-1-27}
{\cal P}=\frac{\sigma\tilde\kappa e^{-b\left(\phi-\phi_{\infty}\right)}e^{-\alpha\phi}}{\left[1-\sigma\tilde\kappa^2 e^{-2b\left(\phi-\phi_{\infty}\right)}\right]^{1/2}},
\ee
where Eq. (\ref{17-1-4}) has been used and we put $\tilde{\kappa}=\kappa e^{-b\phi_{\infty}}$. It is easy to see that the quantities $\phi-\phi_{\infty}$ and $\tilde{\kappa}$ are invariants of a dilatation transformation, because the map $\phi_{\infty}\rightarrow \phi_{\infty}+\Lambda$ holds for $\phi$-asymptotics just as for any other value in this field. In fact, we mean the invariant parameter $\tilde{\kappa}$ in Eq. (\ref{17-1-12}); this parameter coincides with $\kappa$ in the case $\phi_\infty=0$, see Eqs. (\ref{17-1-12}) and (\ref{17-1-27}).    
Thus, Eq. (\ref{17-1-12}) is consistent with Eq. (\ref{17-1-11-1}), and the ratio $q/m$  is a quantity invariant under the action of dilatation transformation.  

The results of this article include the discovery of a new integral of motion (\ref{17-1-17}) of energy type for the dynamics of a test particle on a stationary background. In the case of a spherically symmetric electrostatics we have found potential (\ref{17-1-21}) for the considered dynamics; 
the explicit form of the corresponding potential energy is given by Eqs. (\ref{17-1-24}), (\ref{17-1-25}) and (\ref{17-1-26}). Finally, this  dynamics is solved in quadratures, see Eq. (\ref{17-1-22}). 

The study of string theory and other Grand unification theories led to the discovery of dynamical systems, which include DME as an essential part.
Some of these field theories are representable in the form of a matrix-valued generalizations of the classical gravity (see \cite{OK-1-0}--\cite{OK-3-2} for details) and allow research methods developed in General Relativity (see,  for example, \cite{Ehlers}--\cite{Mazur}). It seems possible to extend the results obtained for DME in this article on these more complex and realistic systems.

\section*{Acknowledgments} We are grateful to our colleagues for useful discussions of the results presented in this article.

\end{document}